\documentclass[12pt]{article}
\usepackage{amsmath,amssymb,graphicx,authblk}
\usepackage{hyperref}

\date{}
\begin{document}
	
\title{Neutron absorption in supermirror coatings: Effects on shielding}

\author[1]{Rodion Kolevatov \thanks{rodion.kolevatov@ife.no}}

\author[2]{Christian Schanzer \thanks{christian.schanzer@swissneutronics.ch}}

\author[2,3]{Peter B\"oni\thanks{pboeni@frm2.tum.de}}

\affil[1]{NMat department, Institutt for Energiteknikk, Instituttveien 18; Postboks 40, 2027 Kjeller, Norway}
\affil[2]{SwissNeutronics AG, Br\"uhlstrasse 28, CH-5313 Klingnau, Switzerland}
\affil[3]{Physik-Department E21, Technische Universit\"at M\"unchen, James-Franck-Str. 1, 85748 Garching, Germany}
		
\maketitle		

\begin{abstract}
The quantification of the dose rate and the composition of the dose in the vicinity of supermirror coated guides is essential for designing shielding for  beamlines transporting neutrons with a high flux. We present results on a calculation of radiative neutron absorption in Ni/Ti and NiMo/Ti supermirror coatings which leads to the emission of high energy gamma rays. A simple parameterization of the absorption probability in the coating materials per incident neutron is given as a function of momentum transfer at reflection. 
\end{abstract}

\section{Introduction}

Supermirrors for the reflection of neutrons \cite{Mezei} are made of multiple alternating layers of Ni and Ti with varying thicknesses and a strong neutron optical contrast. Using such multilayer coatings in neutron guides allows to significantly improve the transport of neutrons compared to guides with pure Ni coating due to the large increase in the angle of total reflection $\theta_c$ at the guide walls by $m$-times the critical angle of reflection $\theta_c^{Ni}$ of Ni, i.e. $\theta_c = m \theta_c^{Ni}$. $m$-values up to $m = 8$ have been reported \cite{Schanzer}.

\pagebreak

During the transport in the guide, neutrons are lost due to the finite reflectivity of the reflecting coating. The major loss mechanisms are caused by absorption, diffuse scattering, and finite transmission through the coating. While the transmitted or scattered neutrons are absorbed in the guide substrate or surrounding absorber releasing gamma radiation of comparatively low energy, absorption of neutrons in the coating materials releases high-energy photons with energies up to $9\div10$~MeV, a typical nucleon binding energy. At the European Spallation Source (ESS) which is currently under construction in Lund, Sweden, the peak flux of transported cold and thermal neutrons is expected to be $1\div 2$ orders of magnitude higher than at the presently operating neutron facilities \cite{ESSTDR}. At the ILL, the brightness of the thermal neutron source is comparable to the time averaged brightness of the ESS. The absorption rate of neutrons in the guide coating will be proportional to the flux. At the long instruments at ESS, gamma radiation will dominate the dose rate beyond the line of sight to the moderator, which typically constitutes more than 2/3 of the total instrument length. Therefore, during the  course of designing shielding for the guides transporting high neutron fluxes, absorption of neutrons in the coating must be carefully addressed. 

To the best of the authors' knowledge out of the existing Monte-Carlo transport codes used for the purpose of shielding calculations, there are currently only two,  PHITS \cite{doi:10.1080/00223131.2013.814553} and MCNP \cite{waters2007mcnpx} which partly implement supermirror physics\footnote{ While the supermirror option is included in the standard PHITS distribution, in MCNP there is an alternative either to use custom libraries which have limited availability for users due to licensing or to interface the transport simulation with neutron ray-tracing software \cite{klinkby2013interfacing}.}. As the coherent scattering of thermal neutrons in the multilayers is not accessible in the transport Monte-Carlo approach, the aforementioned codes only implement parameterizations of the specular reflection probability of the supermirror coatings; the non-reflected neutrons are transported unaffected further beyond the reflecting surfaces. This straightforward approach, however, has serious drawbacks as will be discussed during the course of the paper, which could severely underestimate the actual neutron capture rate in the coatings.

While the reflectivity of supermirror coatings has been extensively studied for a long time also with account of the fabrication process (see e.g. \cite{Schanzer,DW,Pleshanov}), absorption of neutrons by supermirror coatings received much less attention so far. In this note we calculate absorption in the multilayer with layer sequences used in real fabrication. For the calculation, a matrix formalism \cite{Yamada} is used which is briefly described in sect.~\ref{sect:formalism}, followed by a discussion of approaches to account for the interface roughness in multilayers in sect.~\ref{sect:roughness}. In section~\ref{sect:pars}, parameters used in the numerical calculation are outlined. The results on calculations of neutron absorption in supermirrors are presented and discussed in sect.~\ref{sect:results}. We provide guidelines for shielding applications in sect.~\ref{sect:guidelines}, where a parameterization for a absorption probability in the supermirror coating materials is proposed and compared to a calculation. The paper is concluded in sect.~\ref{sect:conclusions}

\section{Neutron waves in a multilayer}
\label{sect:formalism}
\subsection{Matrix formalism}

The wave function $\Psi(\vec r)$ of a neutron propagating in a medium is a solution of the Schr\"odinger equation including an effective potential \cite{Sears,Sears2}:

\begin{equation}
\left\{ \Delta + k^2 + 4 \pi \rho F \right\} \Psi(\vec r) =0, \label{eq-Schr}
\end{equation}
where $k$ is the wave number of the propagating neutron wave in vacuum, $\rho$ is the number density of atoms in the medium and $F\equiv F' + i F''$ is the neutron elastic scattering amplitude per atom taken at zero scattering angle\footnote{Commonly referred to as ``forward scattering amplitude'' in quantum scattering theory.}. The local field corrections are of the order of $10^{-4}$ and may be neglected in the real part of $F$  which is expressed by the bound coherent scattering length $b_c$. They are, however, essential in the imaginary part of the effective potential which in addition to the imaginary part of the bound coherent scattering length $b_c''$ receives contributions dependent on bound coherent and incoherent scattering lengths, $b_c$ and $b_i$ respectively~\cite{Sears2}:
\begin{eqnarray}
F'&  = & -b_c \\
F''& = &  b_c''+k b_{i}^2 +  k b_c^2 \cdot\alpha(k). \label{eq:ImF}
\end{eqnarray}
The scattering lengths are tabular values and are related to the  coherent $\sigma_c$ and incoherent $\sigma_i$ parts of the elastic scattering cross section and absorption cross section $\sigma_a$ as \cite{XSSEARS}:
\begin{eqnarray}
\sigma_c = 4\pi |b_c|^2, \quad \sigma_i = 4\pi |b_i|^2,\quad \sigma_a =  \frac{4 \pi}{k} b_c''.
\end{eqnarray}
The interference factor $\alpha(k)$ entering imaginary part of the scattering amplitude $F''$, eq.~\eqref{eq:ImF}, emerges due to interference of the neutron waves scattered by different atoms \cite{Sears} and also has a pronounced wavelength dependence. It is expressed via the integral of the static structure factor $S({\bf q})$ which for the case of homogeneous isotropic system depends only on the absolute value of its argument~\cite{Sears}:
\begin{equation}
\alpha(k)  = \frac{1}{2 k^2} \int\limits_0^{2 k} S(q) q dq. \label{eq:intfactor}
\end{equation}
For the amorphous metals which constitute layers of the supermirror coatings the structure factor is typically close to zero for $q\lesssim 2$~\AA$^{-1}$ and oscillates around unity for $q \gtrsim 5$~\AA$^{-1}$~\cite{structfact}. This typical behavior suggests $\alpha(k)\approx 0$ for $k\lesssim 1$~\AA$^{-1}$ ($\lambda\gtrsim 6.3$~\AA) and $\alpha(k)\approx 1$ for $k\gtrsim 5$~\AA$^{-1}$ ($\lambda\lesssim 1.25$~\AA). The transition between the two regimes of the diffuse scattering occurs at neutron wavelengths $\lambda_{\rm tr} \approx 2\div 4$~\AA.
 Qualitatively this can be understood in the following way.
When the wavenumber of the neutrons is substantially larger than typical inverse distance between the atoms so that the neutrons scatter off the individual atoms, the interference between the scattered waves is absent and the integral \eqref{eq:intfactor} reaches values close to unity. On the contrary, with a small wave number the neutron wavelength substantially exceeds the typical interatomic distance and it scatters off an ensemble of atoms. The interference factor \eqref{eq:intfactor} is then small and the diffuse scattering receives contributions only from the incoherent part of the scattering cross section.

The optical theorem relates the imaginary part of the forward scattering amplitude to the neutron total interaction cross section in the medium,
\begin{equation}
\sigma_t = \frac{4 \pi}{k} F''.
\end{equation} 
The cross section $\sigma_t$ describes attenuation of the incident beam in the medium\footnote{Transmissivity of a material slab of thickness $d$ and atomic concentration $\rho$ is given by $T=\exp[-\rho \sigma_t d/{n'}]$, where $n'\approx 1 - \cfrac{2 \pi \rho b_c}{k^2}$ is the real part of refraction index.}. With the imaginary part of the amplitude $F''$ given by \eqref{eq:ImF} it receives contributions from both absorption and  {diffuse scattering}, as it should. 
\begin{equation}
\sigma_t(k) = \sigma_a(k) +\sigma_d(k)   =  \sigma_a (k) k_0/k + \alpha(k) \sigma_c +\sigma_i .
\end{equation}
While the coherent  and the incoherent scattering cross sections have a weak energy dependence in the thermal and cold neutron energy range, the absorption cross section is inversely proportional to the neutron velocity $v$ (or wave number $k = m_n v /\hbar $) which must be taken into account explicitly. The coherent part of the elastic scattering cross section enters the diffuse scattering contribution scaled by the wavelength dependent interference factor $\alpha(k)$ as described above.

\begin{figure}[!h]
	\centering
	\includegraphics[width=0.7\hsize]{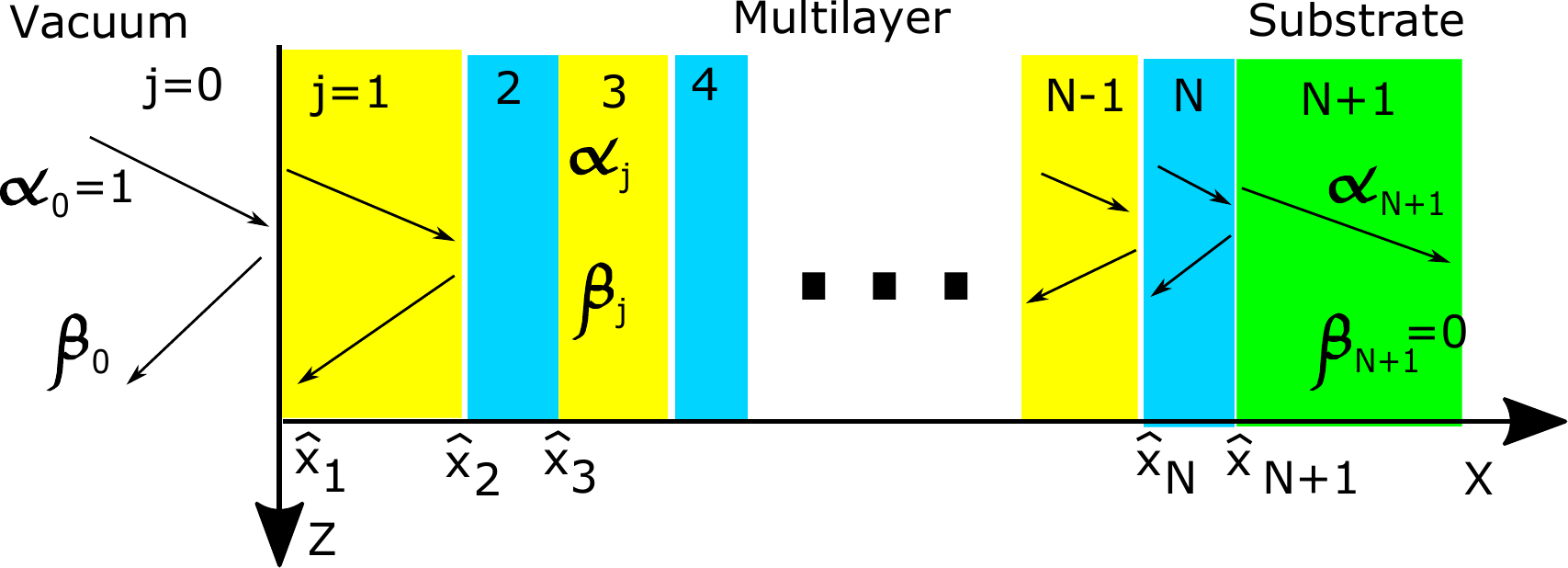}
	\caption{Schematic drawing of the neutron wave incident on the multilayer. Notations are explained in the text. The coefficients $\alpha_0 = 1$ and $\beta_0$ designates the amplitude of the incident and reflected neutron wave.}
	\label{fig:setup}
\end{figure}

Now consider a neutron incident from vacuum  under a small glancing angle on a multilayer structure on a substrate aligned in the $ZY$ plane. In each layer of the multilayer and in the substrate, the wave function of the neutrons satisfies the Schr\"odinger equation Eq.~\eqref{eq-Schr} with the boundary conditions: i) the wave function itself and ii) its derivative are continuous functions at all interfaces. Condition i) ensures conservation of the neutron momentum projection in the $ZY$ plane, which allows factorizing out the lateral part of the wave function:

\begin{equation}
\Psi(x,z) = e^{i k_z z} \psi (x).
\end{equation}
The transverse part of the neutron wave function $\psi(x)$ satisfies the 1-dimensional equation:
\begin{equation}
\left\{ \cfrac{d^2}{dx^2} + k_\perp^2 - 4\pi \rho b_c + i \frac{2 \pi }{\lambda}  \left(\Sigma_d(\lambda) + \Sigma_a(\lambda_{0})\cfrac{\lambda}{\lambda_{0}}\right) \right\} \psi(x) =0, \label{eq-Schr-1D}
\end{equation}
where wavenumber and velocity of the neutrons are expressed in terms of the wavelength $\lambda = 2\pi /k$. In Eq.~\eqref{eq-Schr-1D} the macroscopic cross sections for diffuse elastic scattering and absorption were introduced, $\Sigma_{d/a} (\lambda) = \rho \sigma_{d/a}$.

The solution of \eqref{eq-Schr-1D} for a given layer $j$,  $\psi_j(x)$,  can be written as a sum of incident and reflected waves:
\begin{equation}
\psi_j(x) = \alpha_j e^{i k_{j,\perp} (x - \hat x_j)} +\beta_j e^{-i k_{j,\perp} (x - \hat x_j)},
\label{eq:1Dwavefun}
\end{equation}
where $\hat x_j$ is the coordinate of the interface boundary between layers $j-1$ and $j$, and we assume that the layers are numbered in the direction of propagation of the incident neutron wave starting from $j=0$ (vacuum) and up to $j=N+1$ (substrate) as it is depicted in Fig.~\ref{fig:setup}. By convention, the complex momentum $k_{j,\perp}$ is a solution of the equation
\begin{equation}
k_{j,\perp}^2 =  k_\perp^2 - 4\pi \rho b_c + i \cfrac{2 \pi }{\lambda} \left(\Sigma_d (\lambda) + \Sigma_a(\lambda_{0})\cfrac{\lambda}{\lambda_{0}}\right) \label{eq:complexQ}
\end{equation} with a positive real part. Its imaginary part vanishes in vacuum where the absorption and scattering cross sections are identically zero, so $k_{0,\perp} = k_\perp$.

Writing explicitly the continuity relations for the wave function and its derivative at the interfaces one arrives at the matrix formalism \cite{Yamada,MatrixFormalism}. Pairs of coefficients $(\alpha, \beta)$ in the adjacent layers $j$ and $j-1$ are related via the matrix $T_j$ that is fully determined by the composition and the thickness of the layers:

\begin{equation}
\begin{array}{l}
\displaystyle
\left( \begin{array}{c}
\alpha_j \\ \beta_j
\end{array}\right) = T_j  \left( \begin{array}{c}
\alpha_{j-1} \\ \beta_{j-1}
\end{array}\right);
\\[0.5cm]
\displaystyle
T_j = \left(\begin{array}{l l }
 \frac{k_{j,\perp}+k_{j-1,\perp}}{2 k_{j,\perp}} e^{i k_{j-1,\perp} d_{j-1}} & \frac{k_{j,\perp}-k_{j-1,\perp}}{2 k_{j,\perp}}e^{-i k_{j-1,\perp} d_{j-1}} \\
\frac{k_{j,\perp}-k_{j-1,\perp}}{2 k_{j,\perp}}e^{i k_{j-1,\perp} d_{j-1}} &  \frac{k_{j,\perp}+k_{j-1,\perp}}{2 k_{j,\perp}}e^{-i k_{j-1,\perp} d_{j-1}} 
\end{array}\right).
\end{array} \label{boundary_cond}
\end{equation}
Here $d_{j-1}$ designates the thickness of the layer $j-1$. For layer $j=0$ (vacuum), $d_0$ corresponds to the distance between the plane where phase of the incident/reflected wave equals zero and the surface of the coating. When this plane is chosen to coincide with the coating surface itself, the distance is identically zero, $d_0=0$. This is a typical choice for the calculations which assume zero roughness of the surface.

Without loss of generality, the amplitude of the incident wave can be set to be unity. Amplitudes of the waves reflected and transmitted through the multilayer
can be found from the equation

\begin{equation}
\left( \begin{array}{c}
\alpha_{N+1} \\ 0
\end{array}\right) = T_{N+1}\cdot \ldots\cdot T_1 \left( \begin{array}{c}
1 \\ \beta_{0}
\end{array}\right) \equiv \mathbb{ T} \left( \begin{array}{c}
1 \\ \beta_{0}
\end{array}\right).
\end{equation}
$\mathbb{ T} = T_{N+1}\cdot \ldots\cdot T_1$ is the product of $T$-matrices representing each individual layer. Taking into account explicitly that there is no reflected wave propagating in the substrate ($\beta_{N+1} =0$) one easily finds the amplitude of the  reflected wave $\beta_0$:
\begin{equation}
\beta_0 = -\frac{\mathbb{ T}_{21}}{\mathbb{ T}_{22}}.
\end{equation}
Once $\beta_0$ has been evaluated, amplitudes of the reflected and transmitted wave for every layer can be calculated by successive multiplication of  $(1;\beta_0)$ by matrices $T_j$.

\subsection{Absorption}

Absorption of neutrons in the multilayer can be determined from a change in the quantum mechanical probability current. In 1-dimension it reads according to the definition:

\begin{equation}
J_x = \frac{\hbar}{2 m} \left(\psi^* \frac{\partial}{\partial x}\psi - \psi \frac{\partial}{\partial x}\psi^* \right) = \frac{\hbar}{ m}\,{\rm Im}\left( \psi^* \frac{\partial}{\partial x}\psi\right). \label{probcurrent}
\end{equation}
Continuity of the wave function and its derivative ensure continuity of the probability current at the interface of the adjacent layers. Substituting the neutron wave function into the definition Eq. \eqref{probcurrent} and dropping irrelevant constant factors, the probability current at the outermost and innermost interfaces of layer $j$ becomes:
\begin{eqnarray}
J_{j}(\hat x_j)\!\! &\! \propto\! & \!\! {\rm Im} \left( \left[\alpha^*_j+\beta^*_j\right] \left[i k_{j,\perp} \alpha_j - i k_{j,\perp} \beta_j\vphantom{\beta^*_j}\right]\right), \label{eq:probcurrentA}\\
J_{j}(\hat x_{j+1}) \!\!\! &\! \propto \! &\!\! {\rm Im} \left( \left[\alpha_j e^{-ik_{j,\perp} d_j} +\beta_je^{ik_{j,\perp} d_j}\right]^* \, \left[i k_{j,\perp} \alpha_j e^{ik_{j,\perp} d_j} - i k_{j,\perp} \beta_j e^{-ik_{j,\perp} d_j}\right]\right)\!.\hspace{8mm} \label{eq:probcurrentB}
\end{eqnarray}
Probability currents for the incoming neutron wave and the neutron waves reflected and transmitted by the multilayer are 
\begin{equation}
J_{in} \propto k_\perp |\alpha_0|^2; \quad J_{ref} \propto k_\perp |\beta_0|^2; \quad J_{out} \propto {\rm Re} ( k_{N+1,\perp}) |\alpha_{N+1}|^2.
\end{equation}
Transmission and reflection per one incident neutron commonly referred to as transmission and reflection coefficients are given by the ratios of the corresponding currents:
\begin{equation}
R = \cfrac{J_{ref}}{J_{in}} = \cfrac{|\beta_0|^2}{|\alpha_0|^2};\quad T = \cfrac{J_{out}}{J_{in}} = \cfrac{{\rm Re}(k_{N+1,\perp})}{k_\perp}\cfrac{|\alpha_{N+1}|^2}{|\alpha_0|^2}. \label{eq:RandT}
\end{equation}

The loss in layer $j$ per incident neutron is given by the ratio of the difference of currents at the interfaces of the layers and the probability current of the incoming wave:
\begin{equation}
L_j = \cfrac{J_{j}(\hat x_j)-J_{j}(\hat x_{j+1})}{J_{in}}. \label{eq:Loss}
\end{equation}
Attenuation of the probability current within the layer is either due to absorption or diffuse scattering. Note that we ignore the losses due to diffraction from the grains in the layers of the supermirror that may occur for $\lambda$ smaller than the Bragg edge. For thermal and cold neutron energies, the diffuse scattering is isotropic, so, once the neutron has been scattered, the probability that it will be absorbed in the coating afterwards is negligible. Thus absorption per one incident neutron of a given wavelength and $k_\perp$ in a given layer $j$ by a particular component C is obtained as losses multiplied by the ratio $n^{\rm C}_a(j)$ of macroscopic absorption cross sections of the component $\Sigma_a^{\rm C}$ to the total interaction cross section in the layer material:
\begin{equation}
f^{\rm C}_a (j) = 
\frac{\Sigma_a^{\rm C} (\lambda_{0}) \lambda/\lambda_{0}}{\Sigma_{d}(\lambda) + \Sigma_a (\lambda_{0}) \lambda/\lambda_{0} } \cfrac{J_{j}(\hat x_j)-J_{j}(\hat x_{j+1})}{J_{in}}  \equiv n_a^{\rm C}(j) \cdot \cfrac{J_{j}(\hat x_j)-J_{j}(\hat x_{j+1})}{J_{in}}. \label{eq:acc_abs}
\end{equation}
In Eq. \eqref{eq:acc_abs} the $1/v$ dependence  of the absorption cross sections is accounted for explicitly. $\lambda_{0}=1.798$~\AA\ designates the characteristic wavelength of thermal neutrons which corresponds to the velocity $v= 2200$~m/s.

\section{Interfaces roughness}
\label{sect:roughness}
The reflectivity of supermirror coatings produced according to a specific algorithm for the layer sequence is usually somewhat lower than what follows from the calculation accounting only for elimination of neutrons from the parallel beam due to the diffuse scattering and absorption by individual atoms in the  coating material.
A major source of this reduction is attributed to the roughness of the interfaces layers which leads to destructive interference of the reflected waves. Different approaches may be adopted to model the effects of interface roughness.

\subsection{Debye-Waller factors}
Reduction in the specular reflectivity due to local variations in the position of the layer boundaries can be parameterized by introducing Debye-Waller factors in Eq. \eqref{boundary_cond}. To account for imperfect reflection/transmission due to destructive interference, the amplitudes of the reflected and transmitted waves are adjusted and the following replacements are made \cite{DW}:
\begin{equation}
\alpha_{j-1} \mapsto \alpha_{j-1}; \quad \alpha_j\mapsto \alpha_j \Delta^{\rm t}; \quad \beta_{j-1} \mapsto \beta_{j-1} \Delta^{\rm r}; \quad \beta_{j} \mapsto \beta_{j}, 
\end{equation}
where $1/\Delta^{\rm r} = \exp(- 2 \eta^2 k_\perp^2) <1$ and $1/\Delta^{\rm t}=\exp[-(\eta^2/2)(k_{1,\perp}-k_{2,\perp})^2]<1$ are damping coefficients for the reflected and transmitted waves, respectively. Here it is assumed that the rough interface is approximated by step functions and the ''height'' of each step is randomly distributed according to a Gaussian with standard deviation $\eta$.
Eq. \eqref{boundary_cond} is thus written for the adjusted amplitudes, so the wave function and the probability current are not continuous functions of the coordinate $x$ anymore.

\subsection{Smoothed potential in the intermediate layer}
In another approach adopted for fitting the reflectivity of supermirror coatings it is assumed that the optical potential of the neutrons is smoothly varying at the interfaces \cite{Pleshanov}. The ''valleys'' at the surface of the precedent layer are filled with atoms of the material of the next layer, so the optical potential and the coherent scattering length averaged in the plane have a smooth transition from one layer to another. This gradual variation of the potential decreases the reflection of neutron waves at the interface and increases the transmission. At the same time, the continuity of the wave function and its derivative at the interfaces is preserved. Using this model, a fit of the reflectivity profile for supermirror $m=2.5$ was obtained \cite{Pleshanov}. According to that study, the ``mean roughness'' reaches an unreasonably large value $\sigma \simeq 25$~\AA\ after the first few layers.

The mean roughness of the currently used substrates during mass production does not exceed 1~\AA\ as measured by atom force microscopy \cite{Schanzer}. Moreover, the accuracy in layer thickness using the sputtering technique is better than 0.5\% so one would at least expect the mean roughness at the innermost layers to be similar for the coatings produced with same deposition technique independent on their $m$-value. 
However, as we have verified, describing $m=3$ and $m=6$ supermirror reflectivity curves in the smoothed potential approach requires the ``mean roughnesses'' at the innermost layers of $18$~\AA~and $10$~\AA ~respectively. Note that 10~\AA~is practically one half of the thickness of innermost layers for $m=6$ supermirror ($d \simeq 20$ \AA) which implies essentially no areas of pure material within the layer.

The approach of \cite{Pleshanov} has also other problems as it contradicts the results of several studies by other groups. Namely, in \cite{DW} the effect of smoothed potential was found to be less significant compared to a destructive interference of neutron waves reflected at different places of a rough interface (modeled as stochastic steps on the surface with their heights  distributed according to Gaussian law). In addition, in \cite{maruyama2009effect} it has been established experimentally that reduction in the reflectivity of a supermirror at high momentum transfer is mostly due to diffuse scattering of neutrons at the rough interfaces of the layers which do not satisfy the Bragg condition for reflection, that is those which the neutron wave passes before its reflection takes place. The smoothed potential on the contrary implies essentially no diffuse scattering at the interfaces and the reflectivity reduction is solely due to the increased transmission of the reflecting bilayers.

We thus conclude that the roughness extracted in \cite{Pleshanov} may be a result of attributing all losses to the smooth change of effective potential while neglecting other important aspects, so the approach is not applicable for describing reflectivity profiles of coatings with large $m$-values.

\subsection{Monte-Carlo implementation of the reduction of the reflectivity of supermirrors due to roughness.}
\label{subsec:roughness}

The underlying mechanisms of the supermirror reflectivity reduction due to roughness for the layer interfaces are in general well understood. In the study \cite{maruyama2009effect} both the reflectivity and the angular distribution of the diffuse scattered neutrons were studied for a number of different layer sequences experimentally and theoretically. The observed results were reproduced in a numerical calculation.
According to the study the dominant factor which leads to the reflectivity reduction are the fluctuations in the height of the steps making up the rough interfaces decorrelated in the vertical direction (direction normal to the coating surface).

In our work we have adopted an approach to account for the surface roughness which reminds of that picture. 
We imply that due to the interface roughness, the thickness $d_j$ of a particular layer $j$ traversed by an incident neutron propagating along a straight trajectory is a fluctuating quantity. A rough interface of the layer corresponds to a stochastic change of the interface position along the coating normal with a lateral shift.
The neighboring trajectories thus cross different sequences of effective thicknesses of the layers and acquire dissimilar phase shifts at a same depth. The amplitude of the neutron wave propagating in the layer is a result of the interference of the waves which corresponds to particular trajectories.

The implementation of this model in numerical calculations is straightforward. First, a sequence of interface positions is sampled from normal distributions centered at their nominal values with standard deviations defined by the roughness at the corresponding depths. For each given set $i$ of sampled boundaries $\{\hat x_j^{(i)}\}$ a sequence of the amplitudes of reflected and transmitted waves $\{\alpha_j^{(i)}, \beta_j^{(i)}\}$ is calculated from Eq.~\eqref{boundary_cond} with $d_j = \hat x_{j+1}^{(i)} - \hat x_j^{(i)}$. For a given sampling of the interface positions the continuity of the wave function $\psi_j^{(i)} (x)$ and its derivative is thus respected. 

The resulting wave function $\bar \psi_j (x)$ of a neutron in the layer $j$
which takes into account interference resulting from dissimilar phase shifts of the  sampled wave functions  $\psi_j^{(i)} (x)$ is obtained as their arithmetic mean:
\begin{equation}
\begin{array}{l}
\displaystyle \bar \psi_j (x)  \equiv \frac{1}{N}\sum_{i=1}^{N} \psi_j^{(i)} (x) = \bar \alpha_j e^{i k_{j,\perp} (x - \hat x_j)} + \bar \beta_j e^{-i k_{j,\perp} (x - \hat x_j)};\\ \displaystyle \quad \bar \alpha_j \equiv \frac{1}{N}\sum_{i=1}^N\alpha_j^{(i)} e^{i k_{j,\perp} (\hat x_j-\hat x_j^{(i)})} ; \quad \bar \beta_j \equiv  \frac{1}{N}\sum_{i=1}^N\beta_j^{(i)}e^{-i k_{j,\perp} (\hat x_j-\hat x_j^{(i)})}.
\end{array} \label{eq:1Dwf-average}
\end{equation}
As the sampled values $\hat x_j^{(i)}$ at the interface boundaries are used for calculating the amplitudes according to Eq. \eqref{boundary_cond}, they also enter the phase multipliers of the sampled wave functions instead of the nominal values $\hat x_j$ (see Eq.~\eqref{eq:1Dwavefun}). The exponential factors in the expressions for the average amplitudes  $\bar \alpha_j$ and $\bar \beta_j$ are what remains after having singled out the common phase in the sampled wave functions for the incident and reflected waves, $e^{i k_{j,\perp} (x - \hat x_j)}$ and $e^{-i k_{j,\perp} (x - \hat x_j)}$ respectively.

Strictly speaking, the expression  Eq.~\eqref{eq:1Dwf-average}, gives the average wave function $\bar \psi_j (x)$ for the argument which  for any of the sampled sets of interface positions belongs to the layer $j$: $x\in[ \max\limits_i \hat x^{(i)}_j, \min\limits_i \hat x_{j+1}^{(i)}]$ . Constructing an averaged wave function which would explicitly preserve continuity also close to the design values of the interfaces $\hat x_j$  (or $\hat x_{j+1}$) of the layer $j$ requires replacing the corresponding sampled wave functions in \eqref{eq:1Dwf-average} by the wave function $\psi^{(i)}_{j-1}(x)$   ($\psi^{(i)}_{j+1}(x)$),  when  $x<\hat x^{(i)}_{j}$ ($x>\hat x^{(i)}_{j+1}$). 
While within the layer a change in the current is fully characterized by exponential attenuation in the incident and reflected waves, close to the nominal position of the interface, for the argument $x\in[ \min\limits_i \hat x^{(i)}_j, \max\limits_i \hat x_{j}^{(i)}]$, the wave function and, correspondingly, the current exhibit a more rapid change as a result of a destructive interference of the sampled wave functions.

In our numerical calculations we extrapolate Eq.~\eqref{eq:1Dwf-average} to the nominal positions of the interfaces, which makes the neutron wave function and the current discontinuous functions of the coordinate $x$ at the layers boundaries. The emerging discontinuity of the current is a probability of elimination of neutrons from the incident beam due to scattering from the interface roughness. For the interface between layers $j-1$ and $j$ this probability reads:
\begin{equation}
S_j = \frac{J_{j-1}(\hat x_j) - J_{j}(\hat x_j)}{J_{in}}. \label{eq:LossRoughness}
\end{equation} 
Currents \eqref{eq:probcurrentA} and \eqref{eq:probcurrentB} entering the loss probabilities per one incident neutron, Eqs. \eqref{eq:Loss} and \eqref{eq:LossRoughness}, the same as reflection probability $R$, Eq.~\eqref{eq:RandT}, are calculated using the averaged values $\bar \alpha_j$ and $\bar \beta_j$.

For the plane where the phase of the incident and reflected plane in the vacuum is zero we choose the nominal position of the outer surface of the coating, $x = \hat x_1$. The amplitude of the reflected wave with account of destructive interference at the interfaces and the outermost coating surface is thus obtained as
\begin{equation}
\bar \beta_0 = \frac{1}{N} \sum_{i=1}^{N} \beta_0{(i)}.
\end{equation}

\section{Parameters used for the calculation}
\label{sect:pars}

\subsection{Layer composition}
The non-polarizing  coatings of the supermirrors available on the market \cite{SNwww,Mirrotronwww,s-dhwww}
represent alternating layers of either Ni or an alloy of Ni and Mo, and Ti. A systematic study of the reflectivity of the NiMo alloys with different composition was performed in \cite{Padiyath}. No difference in the reflectivity was observed between coatings of pure Ni and NiMo alloy with a 9.8\% number concentration of Mo used in fabrication at SwissNeutronics. This indicates a similar value of the coherent scattering length density and corresponds to the absolute values of the atomic concentrations of Ni and Mo in the alloy as indicated in Tab.~\ref{tab:at_conc}.

Throughout the calculation we use sequences of layer thicknesses obtained by means of the formalism by Hayter and Mook \cite{Hayter} which coincide with those used in the commercial production of supermirrors \cite{Schanzer,SNwww}. Values of  neutron cross sections and coherent scattering lengths used throughout the work were taken from \cite{XSSEARS}. The wavelength scaling of the absorption cross section was implemented explicitly (see Eqs. \eqref{eq:complexQ} and \eqref{eq:acc_abs}).

\begin{table}[!ht]
	\centering

\setlength\tabcolsep{0.7mm}  
\begin{tabular}{|l|c|c|c|c|c|}
	\hline
{Coating material} & {$\rho$, atoms/cm$^3$} & {$b_c$,~fm} & {$\Sigma_c$,~cm$^{-1}$} & $\Sigma_i$,~cm$^{-1}$ & {$\Sigma_a$,~cm$^{-1}$}  \\
	\hline	
	Pure Ni \vphantom{$I^{I^I}$} & $0.09121\cdot 10^{24}$ & \phantom{$-$}10.3 & 1.21 & 0.474  & 0.410  \\
	Ni+Mo (9.8\%at.) & $0.09443\cdot 10^{24}$ & \phantom{$-$}9.95  & 1.17 & 0.453 &   0.382(Ni)+0.023(Mo) \\
	 Pure Ti & $0.05679\cdot 10^{24}$  & $-3.44$ & 0.084 & 0.163 &  0.346 \\
	\hline
\end{tabular}

\caption{Atomic concentrations $\rho$, voherent scattering lengths $b_c$, macroscopic coherent, incoherent and absorption cross sections $\Sigma_c$, $\Sigma_i$ and $\Sigma_a$, respectively, for thermal neutrons ($\lambda_0 = 1.798$~\AA) \cite{XSSEARS} of typical coating materials for supermirrors.}
\label{tab:at_conc}
\end{table}

For a momentum transfer at the angle of total reflection of Ni, $q_c^{\rm Ni} = 0.0218$~\AA$^{-1}$, neutrons typically do not penetrate beyond the uppermost layer which has typically a thickness $d_1 = 700$~\AA. In this case the decrease of the reflectivity is, first, due to absorption and diffuse scattering of the neutrons in the top Ni layer and, second, due to the interference effects described above. For a perfectly even surface, the wavelength dependent fraction of non-reflected neutrons, which are absorbed by the coating can be thus calculated as ratio of the macroscopic absorption cross section to the total macroscopic interaction cross section in the uppermost layer (cf. Eq.~\eqref{eq:acc_abs}). This gives

\begin{eqnarray}
n_a^{\rm Ni}({\rm NiMo}) = \cfrac{\Sigma_a^{\rm Ni}(\lambda_0)\frac{\lambda}{\lambda_0}}{\big[\Sigma_a^{\rm Ni}(\lambda_0)+  \Sigma_a^{\rm Mo}(\lambda_0)\big]\frac{\lambda}{\lambda_0}  + \Sigma_{i}^{\rm NiMo}+\alpha(\lambda) \Sigma_{c}^{\rm NiMo}}; \label{eq:naNiNiMo}\\
n_a^{\rm Mo}({\rm NiMo}) = \cfrac{\Sigma_a^{\rm Mo}(\lambda_0)\frac{\lambda}{\lambda_0}}{\big[\Sigma_a^{\rm Ni}(\lambda_0)+  \Sigma_a^{\rm Mo}(\lambda_0)\big]\frac{\lambda}{\lambda_0}  + \Sigma_{i}^{\rm NiMo} + \alpha (\lambda) \Sigma_{c}^{\rm NiMo} }, \label{eq:naMoNiMo}
\end{eqnarray}
for a NiMo alloy and 
\begin{equation}
n_a^{\rm Ni}({\rm Ni}) = \cfrac{\Sigma_a^{\rm Ni}(\lambda_0)\frac{\lambda}{\lambda_0}}{\Sigma_a^{\rm Ni}(\lambda_0) \frac{\lambda}{\lambda_0}  +\Sigma_{i}^{\rm Ni} + \alpha (\lambda)\Sigma_{c}^{\rm Ni}}; \label{eq:naNiNi}
\end{equation}
for pure Ni. The values calculated according to Eqs.~\eqref{eq:naNiNiMo}--\eqref{eq:naNiNi} using the parameters of table~\ref{tab:at_conc} for evaluating the macroscopic cross sections ($\Sigma_{c/i/a} = \rho\cdot \sigma_{c/i/a}$) are given in  table \ref{tab:captfracinf}. While for the wavelength dependent interference factor $\alpha(\lambda)$ at $\lambda=1$~\AA\ and $\lambda=5$~\AA \ one can assume values 1 and 0, respectively, for the calculations at $\lambda =3$~\AA \ we anticipate a value $\alpha =0.5$.  
\begin{table}[!h]
	\centering	
	\begin{tabular}{l|c|c|c}
		$\lambda$, \AA & $n_a^{\rm Ni}$(NiMo) &  $n_a^{\rm Mo}$(NiMo) & $n_a^{\rm Ni}$(Ni) \\
		\hline
		5 & 0.674 & 0.041 & 0.706\\
		3 & 0.372 & 0.022 & 0.388\\
		1 & 0.115 & 0.007 & 0.119
	\end{tabular}
	
	\caption{Absorption probability per non reflected neutron for momentum transfers $q < q_c^{\rm Ni}$ or $q < q_c^{\rm NiMo}$ assuming zero roughness of the surface of the coating.} \label{tab:captfracinf}
\end{table}

\subsection{Interface roughness, accuracy of deposition and resolution of momentum transfer}
The effect of the interface roughness is modeled as described in sect.~\ref{subsec:roughness}. For the standard deviation of the stochastic displacement of the layer interfaces due to roughness the following parameterization is adopted: 
\begin{equation} 
\sigma_j (h) [\text{\AA}] = 2.0[\text{\AA}] + 0.1 \cdot  h[\mu \text{m}].  \label{eq:roughness}
\end{equation}
It is assumed that $\sigma_j$ grows by 0.1~\AA~per $\mu$m of the deposited thickness $h$ of the multilayer as measured from the substrate. The initial roughness value of 2~\AA\ is required to provide a good description of the measured reflectivity curves. It is a factor 2 larger than the typical roughness of the substrate as reported by~\cite{Schanzer} which may indicate a fast growth of the roughness during deposition of the first layers of the coating. The value is, however, still not compatible with the value of 25~\AA\ quoted in \cite{Pleshanov}. 


Deviations of the layer thicknesses from their nominal values due to the limited  precision of the sputtering plant are taken into account as well. The probability currents calculated using amplitudes corrected for the interface roughness are averaged over the thicknesses $d_j$ which are sampled from a Gaussian distribution. The standard deviation corresponds to the precision of the sputtering plant and constitutes 0.5\% of the layer thickness, i.e. $0.005\, d_j$.  

For comparing calculations with measured reflectivity profiles the divergence of the incident neutron beam is incorporated by introducing a distribution for the transverse momentum $q_\perp$ of the neutrons. It is sampled from a Gaussian distribution with a standard deviation equal to 1.5\% of the critical momentum of total reflection for Ni, i.e. $\sigma (q_\perp) = 0.015 \cdot q_c^{\rm Ni} = 3.27\cdot 10^{-4}$~\AA$^{-1}$. The divergence leads to a broadening of the Bragg peaks constituting  the reflectivity and absorption profiles which are otherwise visible (see the black curve for $m=6$ in the bottom panel of Fig.~\ref{fig:refl}). Smoother curves ease perception and parametrization of the data. We have verified that a few hundred simulations provide sufficient accuracy. Results presented in the paper were obtained by a two-stage Monte-Carlo average with 50 random samplings for a combination $(q_\perp, \{d_j\})$
and 1000 samplings of interfaces to account for the roughness leading to a total of 50000 simulations.

\section{Absorption per incident and per non reflected neutron}
\label{sect:results}

\begin{figure}[!htbp]
	\centering
	
	\includegraphics[width=0.8\hsize]{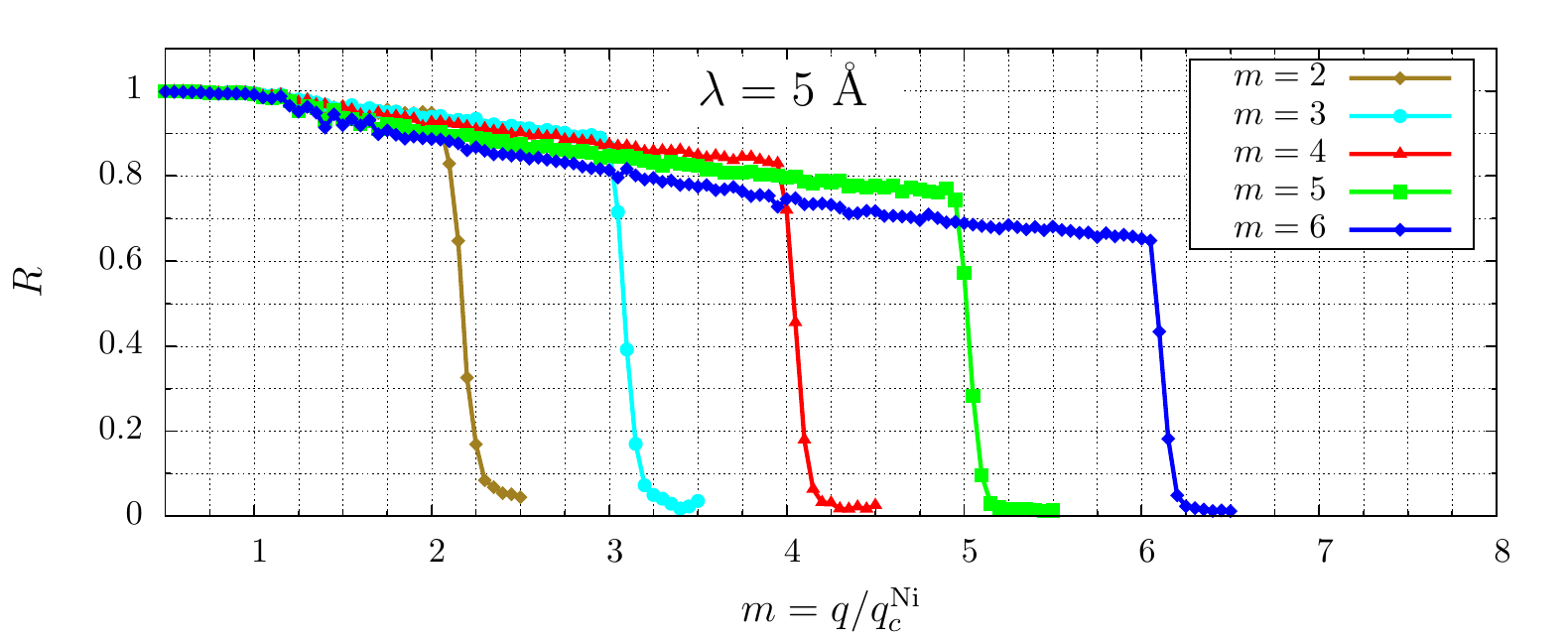}
	
	\includegraphics[width=0.8\hsize]{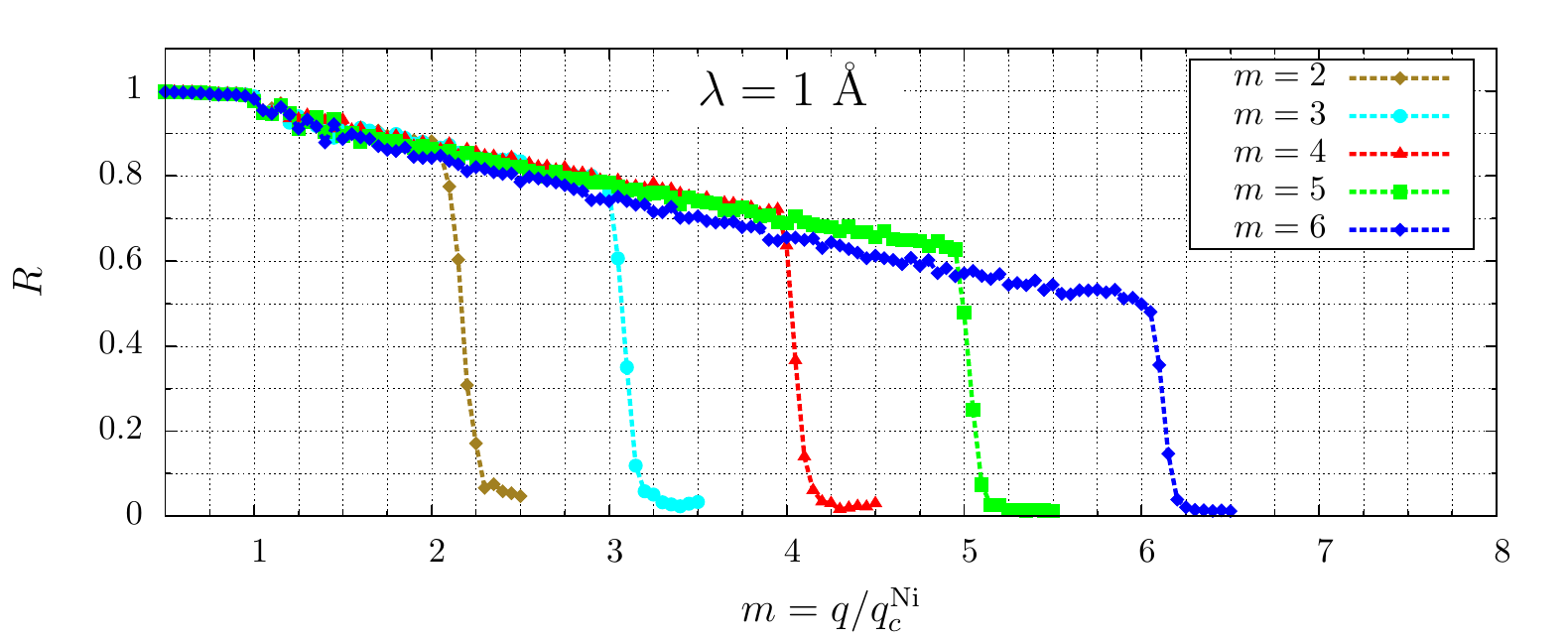}

	\includegraphics[width=0.8\hsize]{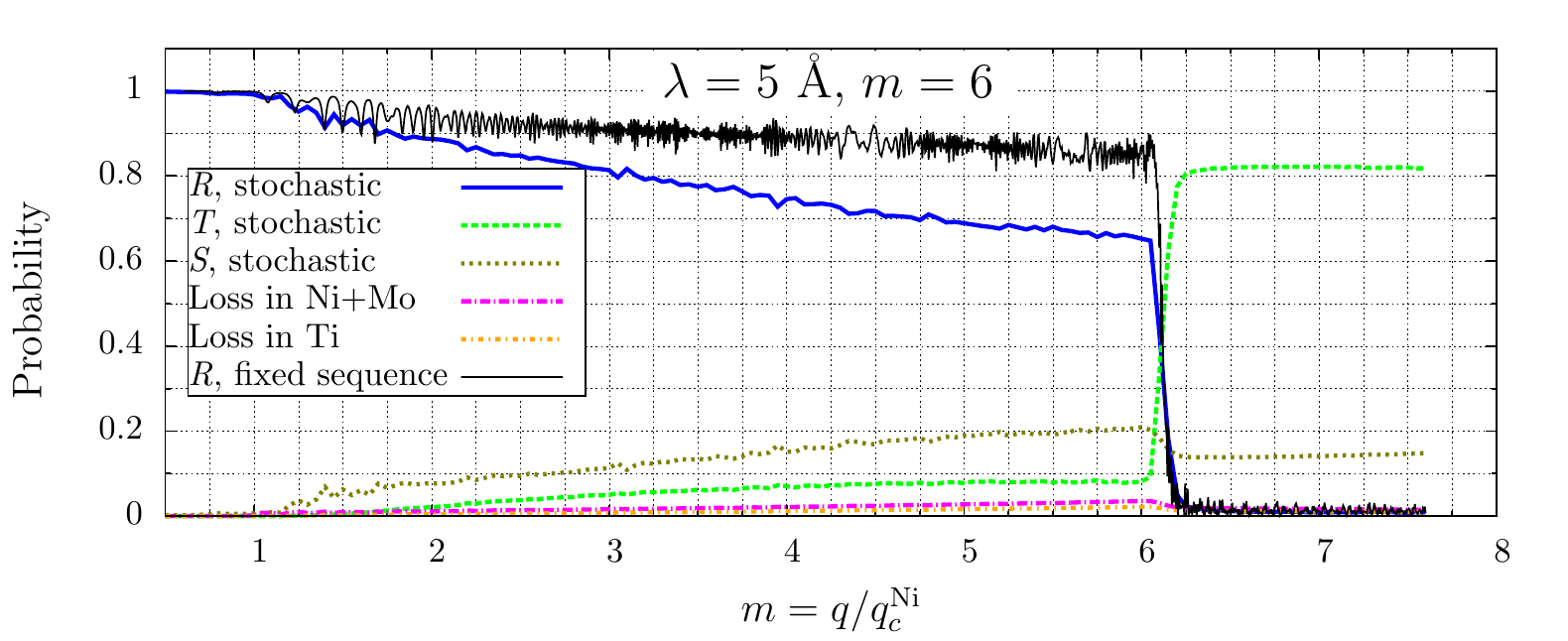}
	
	\caption{Top panel: Calculated reflectivity profiles for $\lambda =5$ \AA\ for the model assuming a fluctuating thickness of the layers to mimic the roughness of the interfaces. Middle panel: Reflectivity for $\lambda=1$ \AA. Bottom panel: Calculated probabilities for reflection $R$ (blue line), transmission $T$ (green line), scattering by interface roughness $S$, and loss (due to absorption and diffuse scattering)  in NiMo and Ti (magenta and orange lines, respectively) for $m=6$ at $\lambda =5$~\AA. The results are compared with a model where the sequence of the layers is fixed to the nominal values assuming zero roughness and zero divergence of the incident beam (black line).}
	\label{fig:refl}
\end{figure}

To benchmark the model, reflectivity profiles for supermirrors $2 \le m \le 6$  were computed for $\lambda = 5$ \AA. The results are plotted in the top panel of Fig.~\ref{fig:refl}. Calculated curves of the supermirror reproduce the measured profiles fairly well \cite{Schanzer}. The calculated reflectivity at smaller wavelengths is found to be lower which is  illustrated in the middle panel of Fig.~\ref{fig:refl}, where reflectivity profiles for $\lambda = 1$~\AA\ neutrons are presented\footnote{Dependence of the reflectivity on the wavelength used in the measurement was mentioned in a study \cite{masalovich2013analysis}}. 

We anticipate the following explanation for the reduction of the reflectivity.
Note that the depth at which the reflection of the neutrons takes place depends solely on the momentum transfer $q_\perp$ at reflection and that for a fixed momentum transfer a shorter wavelength implies a smaller glancing angle. The path length of a neutron in the coating before it has reached the reflection depth thus increases with decreasing wavelength as $\sim 1/\lambda$. The absorption cross section is proportional to the neutron wavelength $\lambda$, so the probability for neutron absorption before it has reached the reflection depth, given in first approximation by a product of absorption cross section times the path length in the coating, is approximately wavelength independent. On the contrary, the diffuse scattering cross section is larger for smaller wavelengths due to the increase of the interference factor. 
Together with the increase of the neutron's path length in the coating for lower glancing angle, this makes probability for removal of neutrons from the incident beam due to the diffuse scattering larger for smaller wavelengths. 
The same attenuation occurs also in the reflected wave. In quantum-mechanical considerations this effect manifests itself as a contribution to the imaginary part of the normal component of the neutron momentum (see Eq.~\eqref{eq:complexQ}) scaling as $1/\lambda$ and being proportional to $\Sigma_d$.

Note that for a given $q$ and fixed $\lambda$, the reflectivity of the supermirror increases with decreasing $m$-value as expected provided $q$ is below the cutoff $m q_c^{\rm Ni}$. We attribute this effect to the lower interface roughness of the layers, where the reflection takes place for coatings with low $m$-value. The effect of roughness is illustrated in the lower panel of Fig.~\ref{fig:refl}. Rough interfaces reduce the reflectivity by up to around 20\% as follows from comparison of the reflectivity profiles for $m=6$ coatings calculated with (blue line) and without fluctuations (black line) in the thicknesses of the coating layers and the positions of the interfaces. This number is in good agreement with the results of experimental study of the diffuse scattering by the rough interfaces in the supermirror \cite{maruyama2009effect}. In the lower panel of Fig.~\ref{fig:refl}, the probabilities of transmission (green line), scattering by the interface roughness (olive) and loss due to diffuse scattering and absorption within the layers (magenta line for NiMo and orange line for Ti) are  plotted. For zero roughness at the interfaces, the conservation of the probability implies that the probabilities for reflection, transmission, and loss within the layers add up to unity.
Fulfillment of this condition was explicitly checked in the calculations.

\begin{figure}[ht]
	
	\centering 
	\includegraphics[width=0.5\hsize]{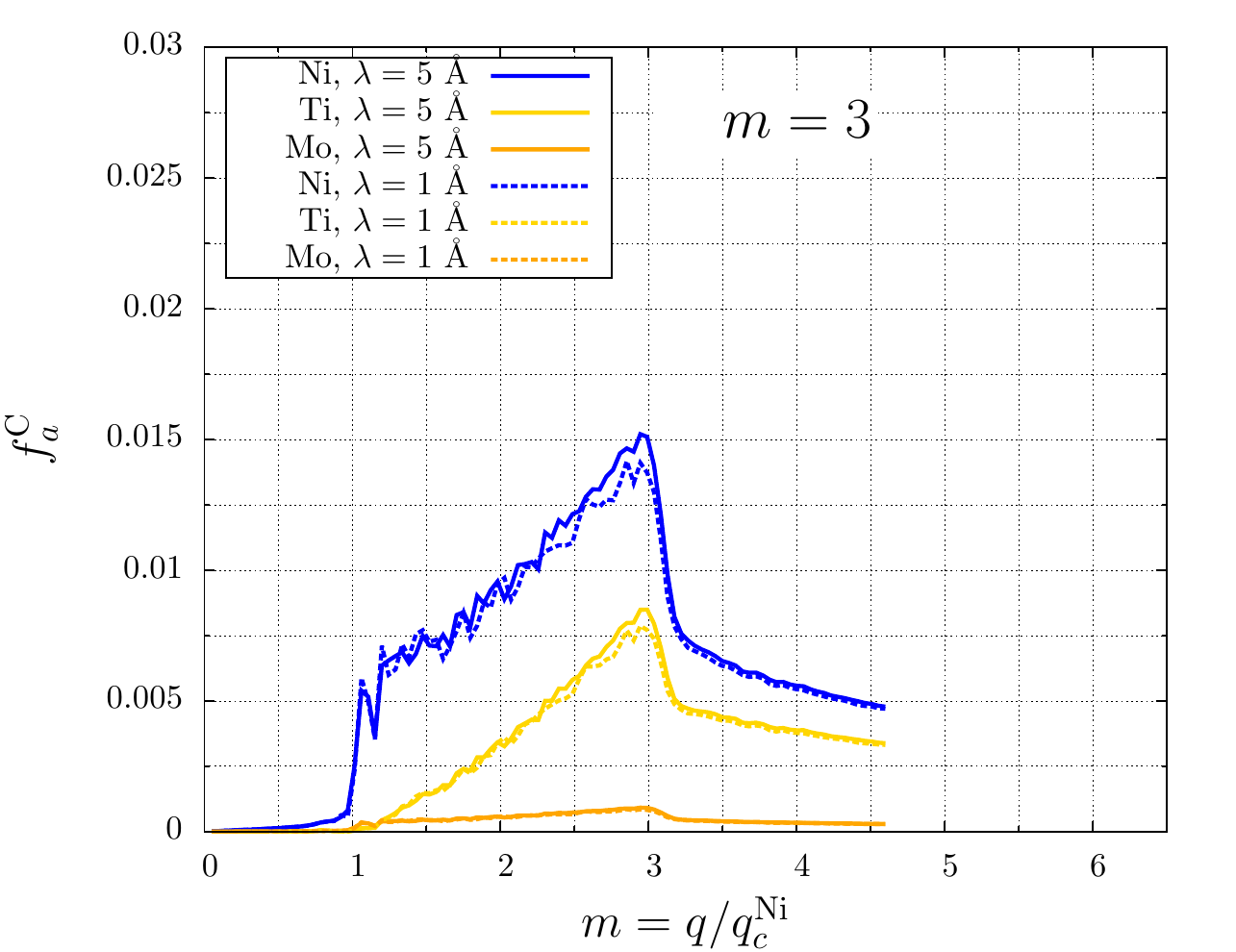}\includegraphics[width=0.5\hsize]{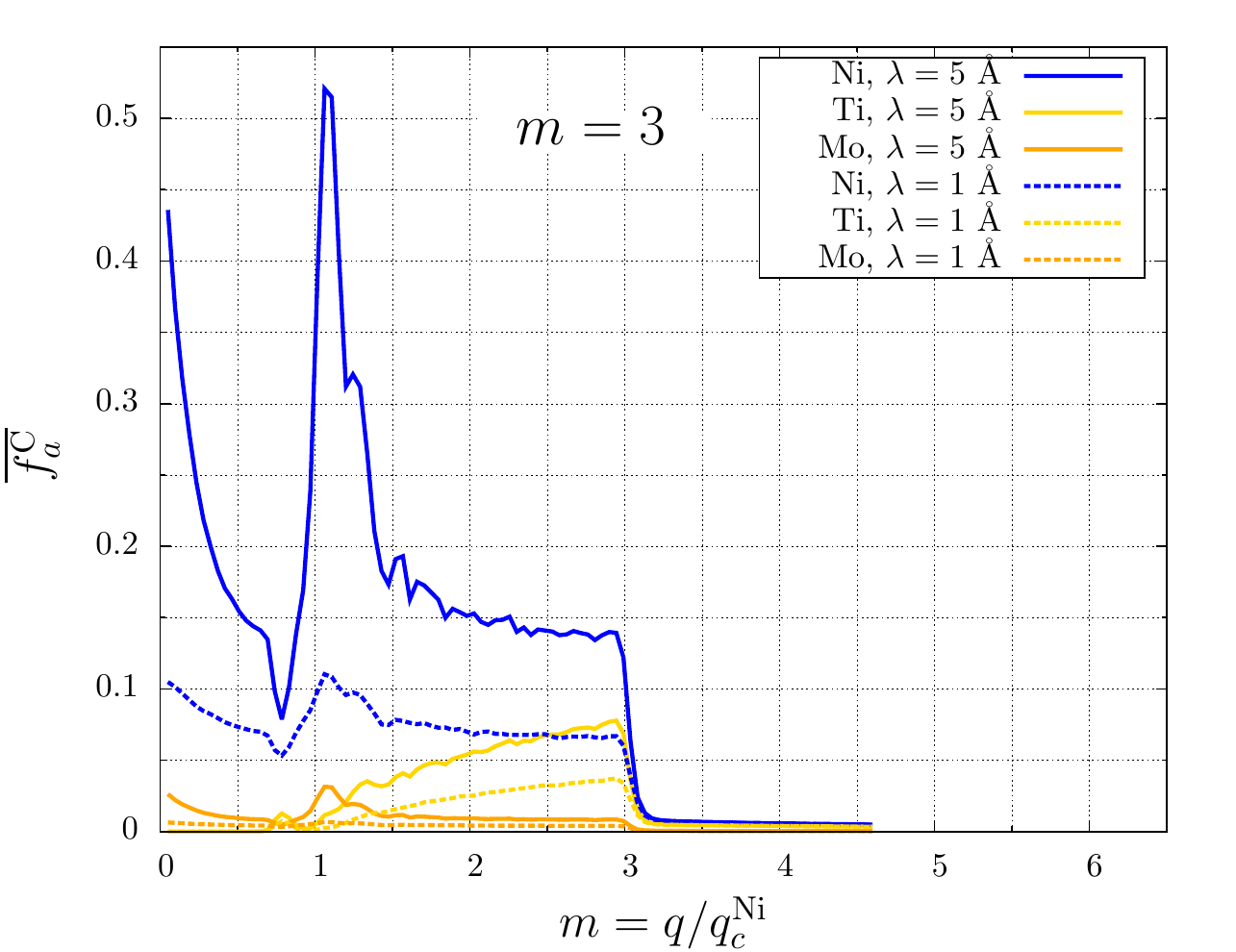}
	
	\includegraphics[width=0.5\hsize]{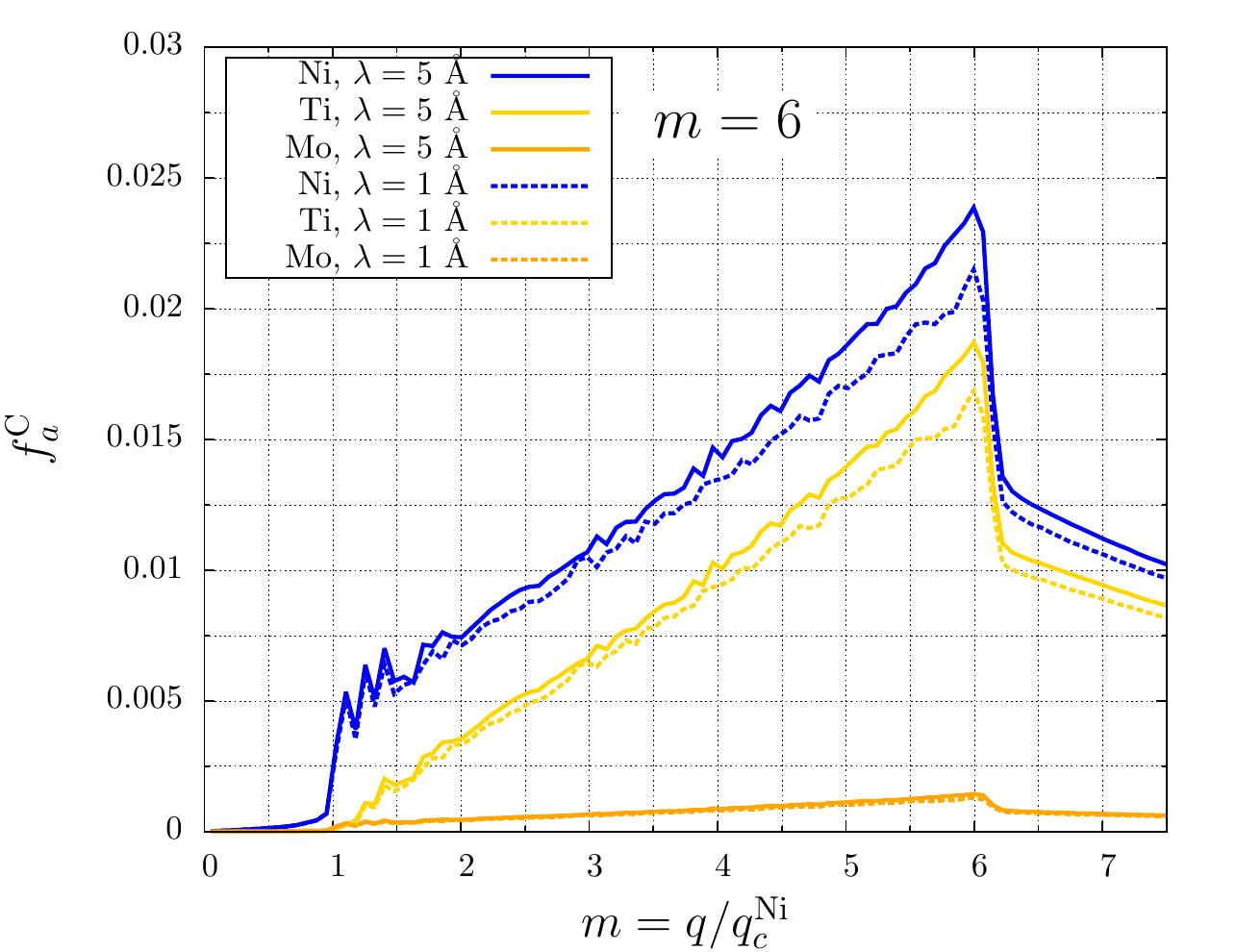}\includegraphics[width=0.5\hsize]{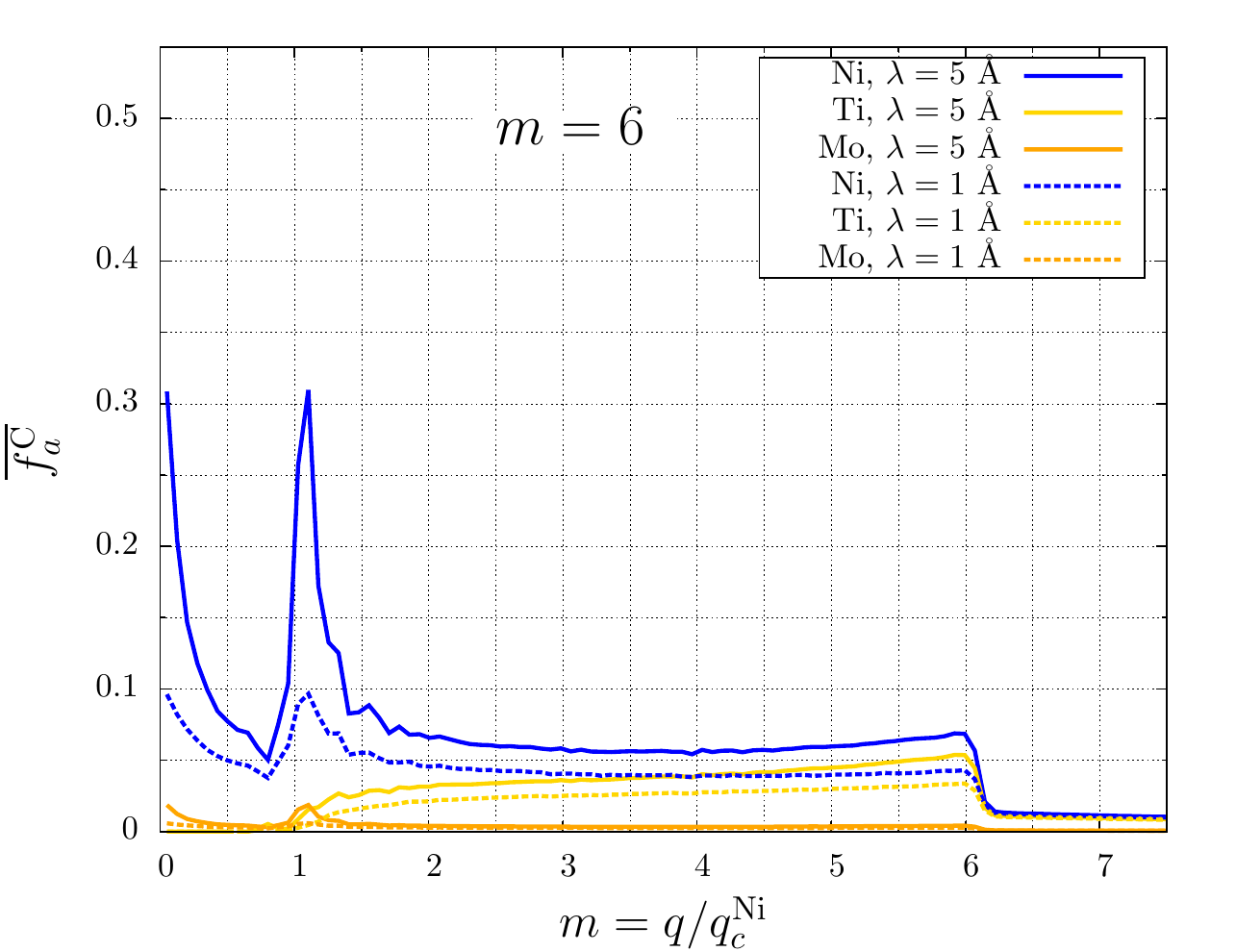}
	
	\caption{Absorption probability per incident neutron (left panels) and per non reflected neutron (right panels) as a function of momentum transfer $q$ for supermirror $m=3$ and $m=6$ for wavelengths $\lambda=1$~\AA \ and  $\lambda=5$~\AA. } 
	\label{fig:capture}	 	
\end{figure}

Next we turn to the absorption of neutrons within the supermirror coating. 
In Fig.~\ref{fig:capture}, the absorption probability per incident neutron and per non-reflected neutron calculated according to Eq.~\eqref{eq:acc_abs} for Ni, Mo, and Ti in supermirror $m = 3$ and $m = 6$ is plotted. Below the cutoff $q \le m q_c^{\rm Ni}$ the absorption per incident neutron exhibits an approximately linear growth, similar for both $m=3$ and $m=6$ coatings indicating that the depth where the neutrons are reflected increases with increasing momentum transfer. This effect is similar for supermirror with different $m$. For $q < m q_c^{\rm Ni}$ neutron absorption occurs for both, the incoming and the reflected wave. In contrast, for $q > m q_c^{\rm Ni}$, the reflected wave is essentially absent. Therefore, at $q =  m q_c^{\rm Ni}$ the absorption probability per incident neutron drops roughly by a factor of 2 as can be clearly seen in the figure. With even further increase of momentum, the absorption probability per incident neutron follows approximately the simple scaling law:
\begin{equation}
f_a^C \propto \cfrac{1}{q}.
\end{equation}
This is a trivial consequence of the fact that the absorption probability is proportional to the path length in the supermirror, which in turn is inversely proportional to the glancing angle and hence to the normal momentum component:
\begin{equation}
 f_a^C \approx \Sigma_a(\lambda) \cfrac{d}{\theta} = \Sigma_a(\lambda_0) \cfrac{\lambda}{\lambda_0}\cfrac {2 \pi d}{\lambda} \cdot \cfrac{1}{q} \ . \label{absbeyond}\end{equation}
It is remarkable that $f_a^C$ is essentially independent of $\lambda$,
which is also true beyond the cutoff $m q_c^{\rm Ni}$ as clearly observed in the left panels of Fig.~\ref{fig:capture}.

The \emph{absorption per non reflected neutron}, $\overline {f_a^C}$, is obtained from $f_a^C$ by dividing $f_a^C$ by $1-R$, where $R$ is the reflectivity. In contrast to the \emph{absorption per incident neutron}, $\overline {f_a^C}$ is a decreasing function of momentum transfer because of the increasing role of the transmission and destructive interference due to the interface roughness. For this reason, $\overline {f_a^C}$ decreases with increasing $m$ for fixed $q$ for  $q\le m q_c^{\rm Ni}$. Because of the reflectivity dependence on $\lambda$, the absorption per non reflected neutron is also a $\lambda$-dependent quantity as clearly seen in the panels on the right hand side of Fig.~\ref{fig:capture}.

 \begin{figure}[!htbp]
 	\centering
 	\includegraphics[width=0.5\hsize]{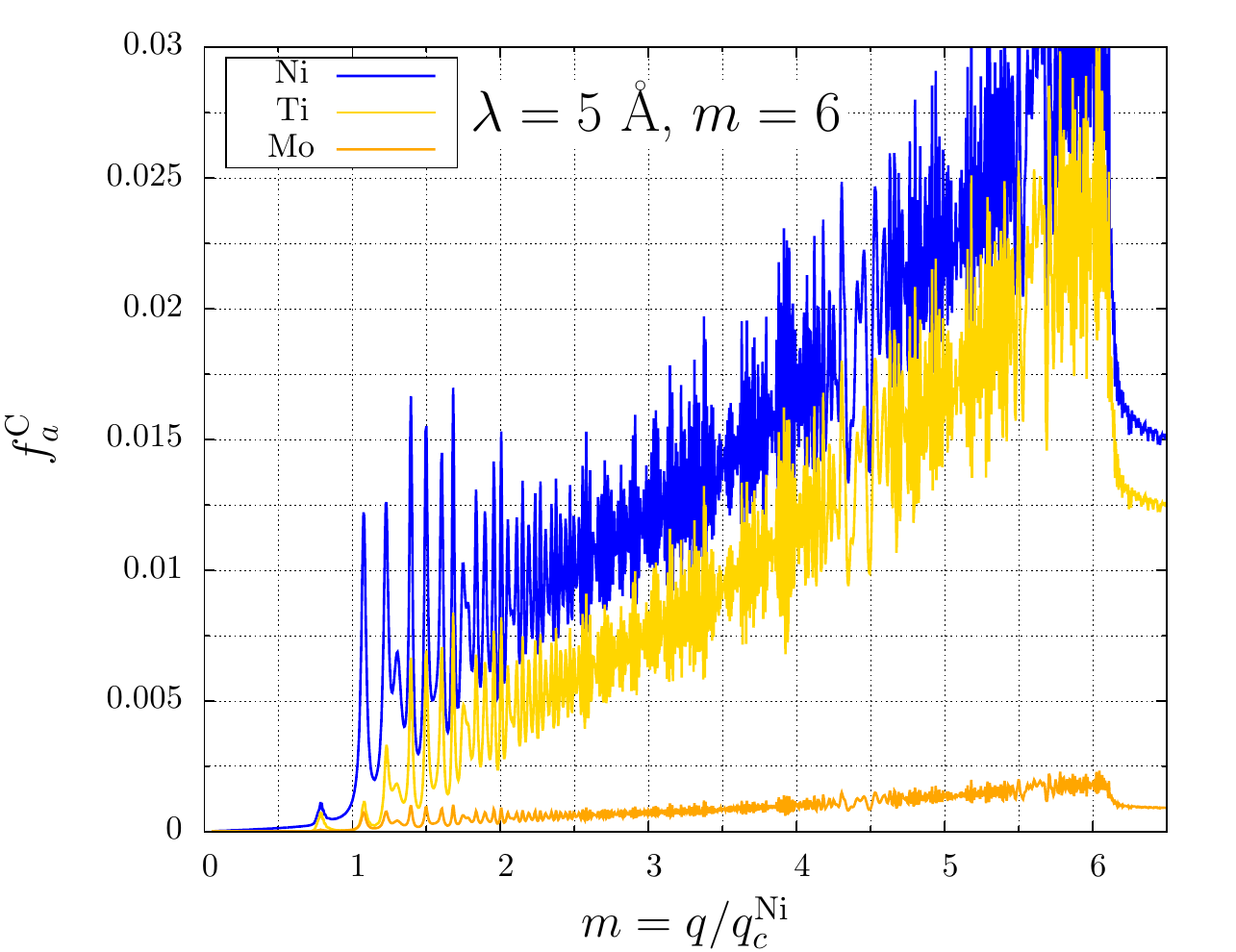}\includegraphics[width=0.5\hsize]{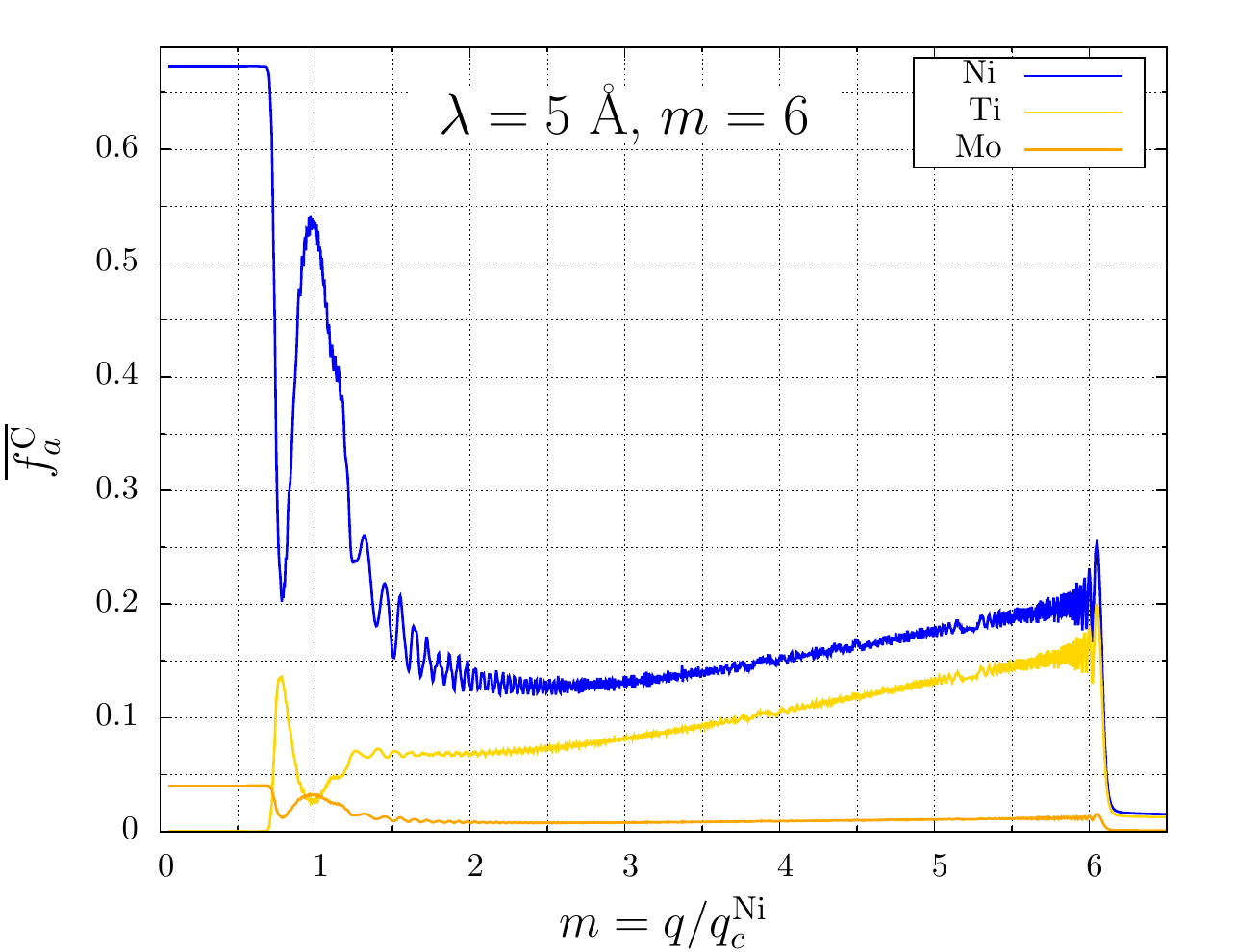}

 	\caption{The absorption probability per incident neutron and per non reflected neutron for supermirror $m=6$ assuming zero roughness and perfect momentum transfer resolution are shown on the left and right hand side, respectively.}
 	\label{fig:m=6zero}	 	 	
 \end{figure}

The linear dependence of $f_a^C$ on $q$ as observed between $q_c^{\rm Ni}$ and  $mq_c^{\rm Ni}$ is essentially independent of the roughness. The independence is seen when comparing Figs.~\ref{fig:capture} and \ref{fig:m=6zero}. Fig.~\ref{fig:m=6zero} shows the absorption per incident and per non-reflected neutron for supermirror $m=6$. For these calculations the thickness of the layers was fixed at the nominal sequence and a roughness equal to zero was assumed. The absorption per incident neutron follows on average the same linear growth as shown in Fig.~\ref{fig:capture}. In contrast, the absorption per a non reflected neutron is larger due to the higher reflectivity when the roughness is set to zero.

For momentum transfers below the critical value $q_c^{\rm Ni}$ there is a peak in $\overline {f_a^C}({\rm Ti})$ at $q \approx 0.75 \cdot q_c^{\rm Ni}$ which is related to the dips in $\overline {f_a^C}({\rm Ni})$ and $\overline {f_a^C}({\rm Mo})$ clearly observed in the calculation assuming zero roughness, Fig.~\ref{fig:m=6zero}. At this $q$, the incident neutron wave spreads beyond the top layer of the supermirror that has a thickness of 700~\AA. Limiting values for absorption per non reflected neutron (see tab.~\ref{tab:captfracinf}) are reached for $q \lesssim 0.6 \cdot q_c^{\rm Ni}$.
In the calculations which take roughness of the coating surface into account presented in Fig.~\ref{fig:capture} a significant fraction of neutrons is not reflected specularly due to the destructive interference arising from the surface roughness. This effect decreases the fraction of not reflected neutrons which are absorbed, $\overline {f_a^C}$, so the limiting values listed in tab.~\ref{tab:captfracinf} are not reached. The reduction in $\overline {f_a^C}({\rm Ni})$ is more pronounced for $m=6$ than for $m=3$ due to the larger roughness of the surface assumed in the model (see Eq.~\eqref{eq:roughness}). With the account of roughness the peak in titanium capture at $q \approx 0.75 \cdot q_c^{\rm Ni}$ is also less pronounced.

\section{Guidelines for gamma shielding of supermirror coated optics}
\label{sect:guidelines}

\subsection{Existing solutions for the transport codes.} 
Specular reflection of cold and thermal neutrons from a multilayer is a coherent process, its description requires explicit account of the wave properties of the incident neutrons. In the Monte-Carlo transport codes which were primarily designed for applications involving neutrons in the fast or thermal energy range, the neutrons are by default treated as  corpuscles which undergoes successive scatterings with the atoms in the medium. Because of this fundamental limitation  modeling of the coherent scattering effects is not accessible. 

To improve the situation, a supermirror option was introduced in PHITS and MCNP transport codes. It implements a specular reflection probability dependent on the momentum transfer at reflection and thus allows for modeling cold and thermal neutron transport along the guides. The non-reflected neutrons, which below the supermirror cutoff typically constitute a small fraction of the incident beam, penetrate the reflecting surface and are transported further in the surrounding materials. 

The implementation, however, is missing an important detail. As it has been mentioned in sect.~\ref{sect:results}, above the critical angle for nickel and below the cutoff, the reflection takes place at a depth where  the layer spacing $d$ satisfies the Bragg's reflection law: $d= \pi/k_\perp$. Hence it is the full incident beam which propagates in the bulk coating up to the reflection depth, not just its non-reflected fraction. Assuming the reflection to take place at the coating surface will thus underestimate significantly the absorption probability.

For a given momentum transfer at reflection (that is fixed wavelength and glancing angle)  the probability for the absorption by the coating could in principle still be obtained in PHITS and MCNP. This can be done by setting the reflecting surface at a depth which corresponds to the momentum transfer under consideration. For example, close to the supermirror cutoff the reflecting surface should be positioned between the coating and the substrate. However for a momentum transfer value different from the one for which this fine tuning was performed, the capture rate obtained in the setup will be obviously incorrect.

\begin{figure}[!htb]
  	\centering	
  	\includegraphics[width=0.5\hsize]{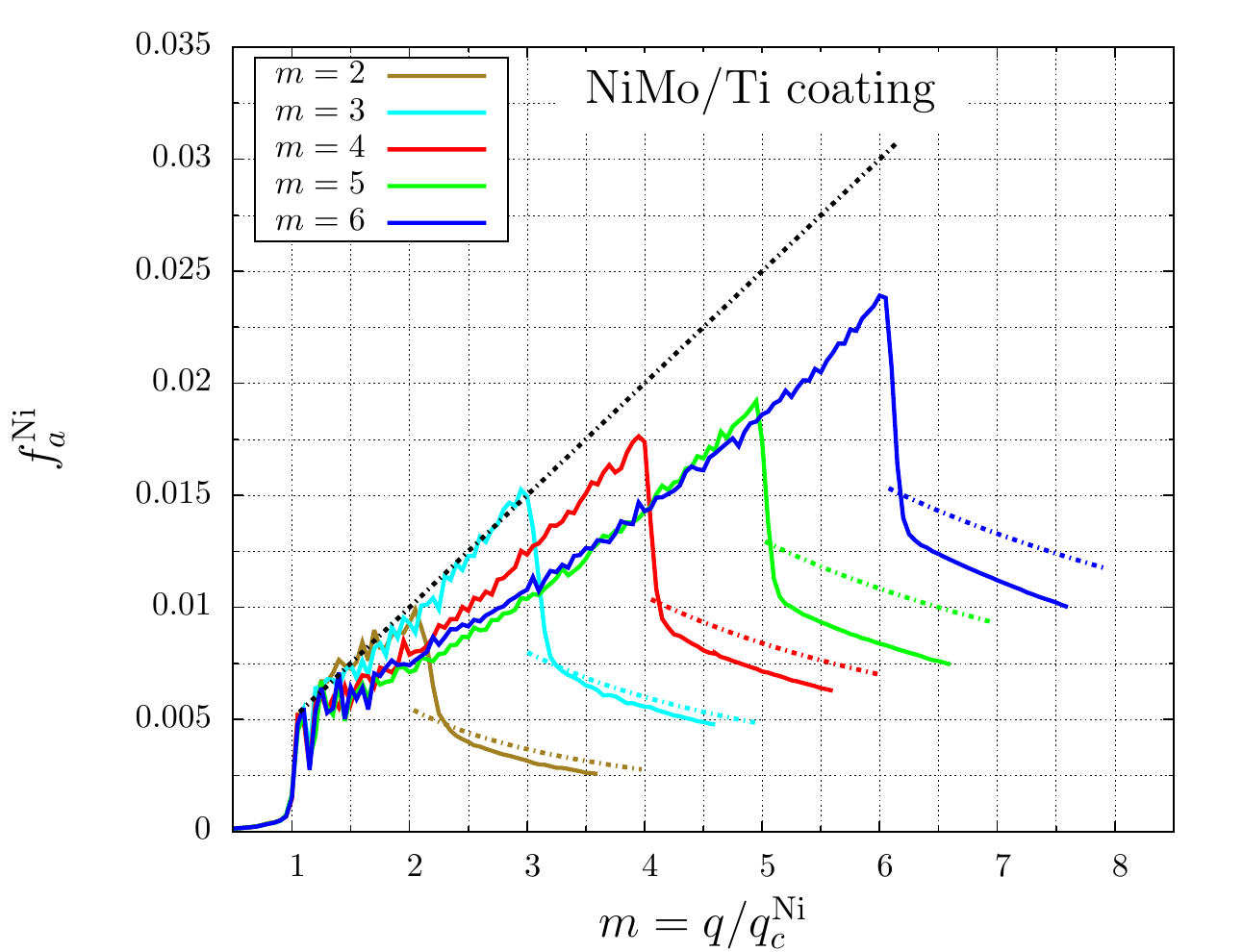}\includegraphics[width=0.5\hsize]{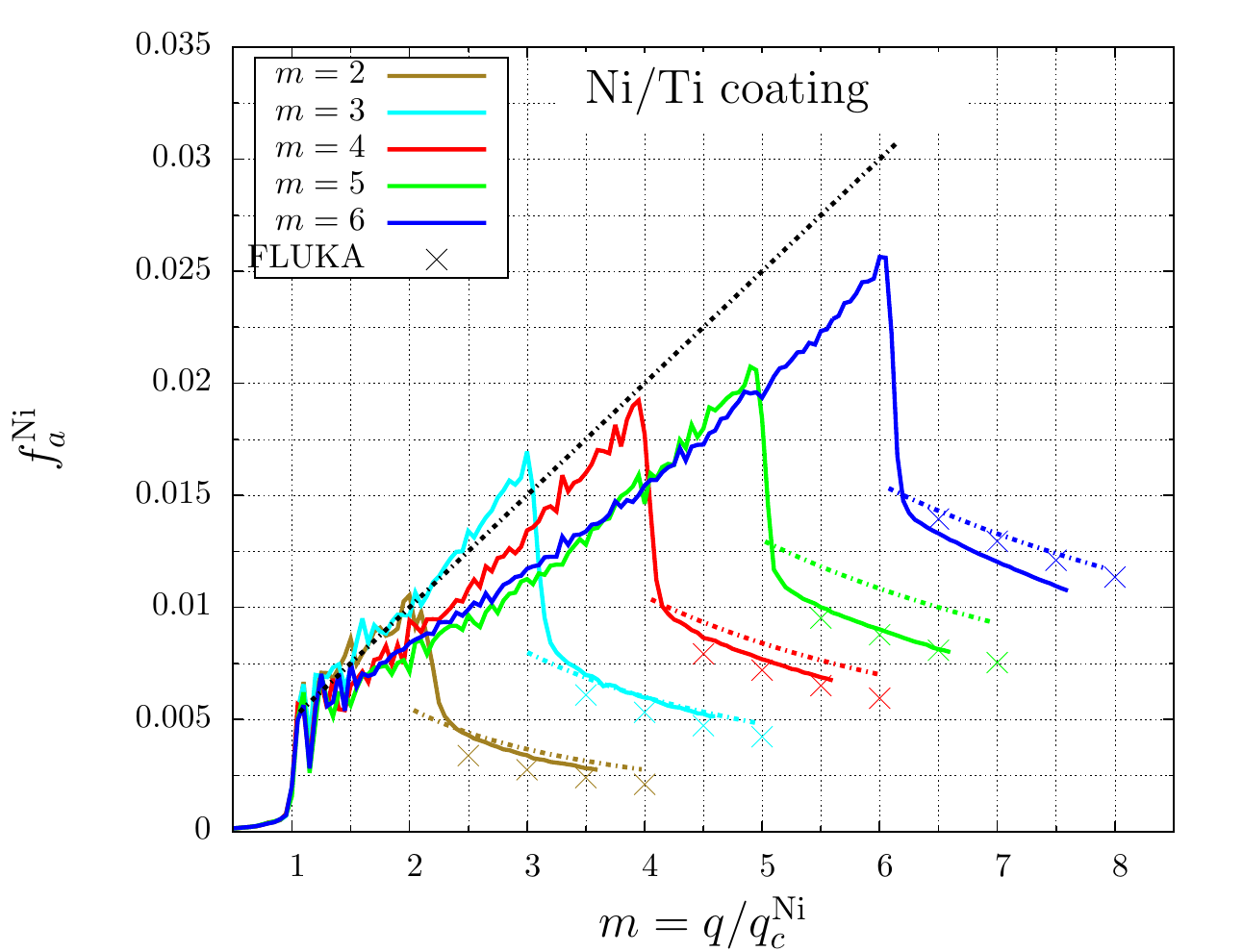}
  	\includegraphics[width=0.5\hsize]{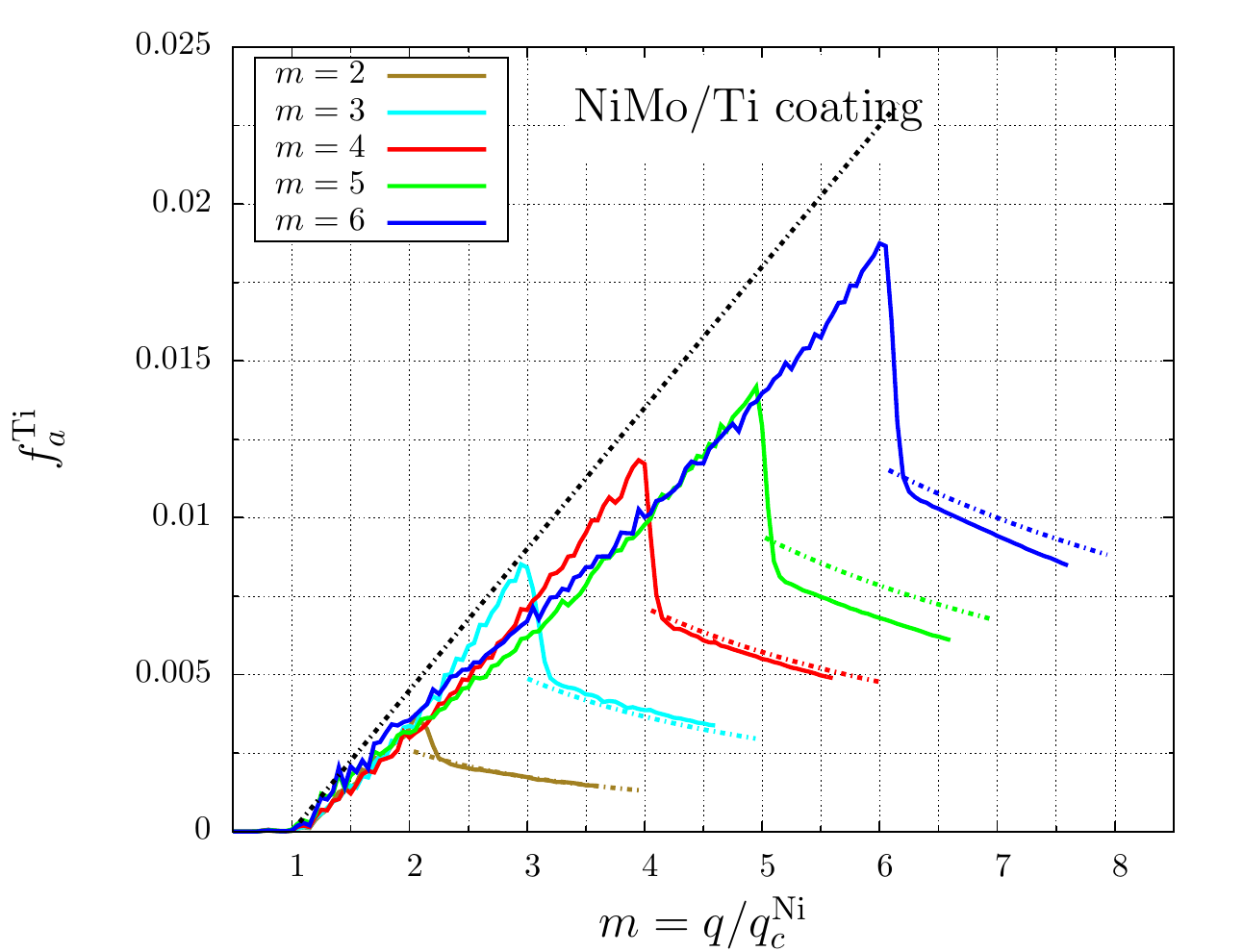}\includegraphics[width=0.5\hsize]{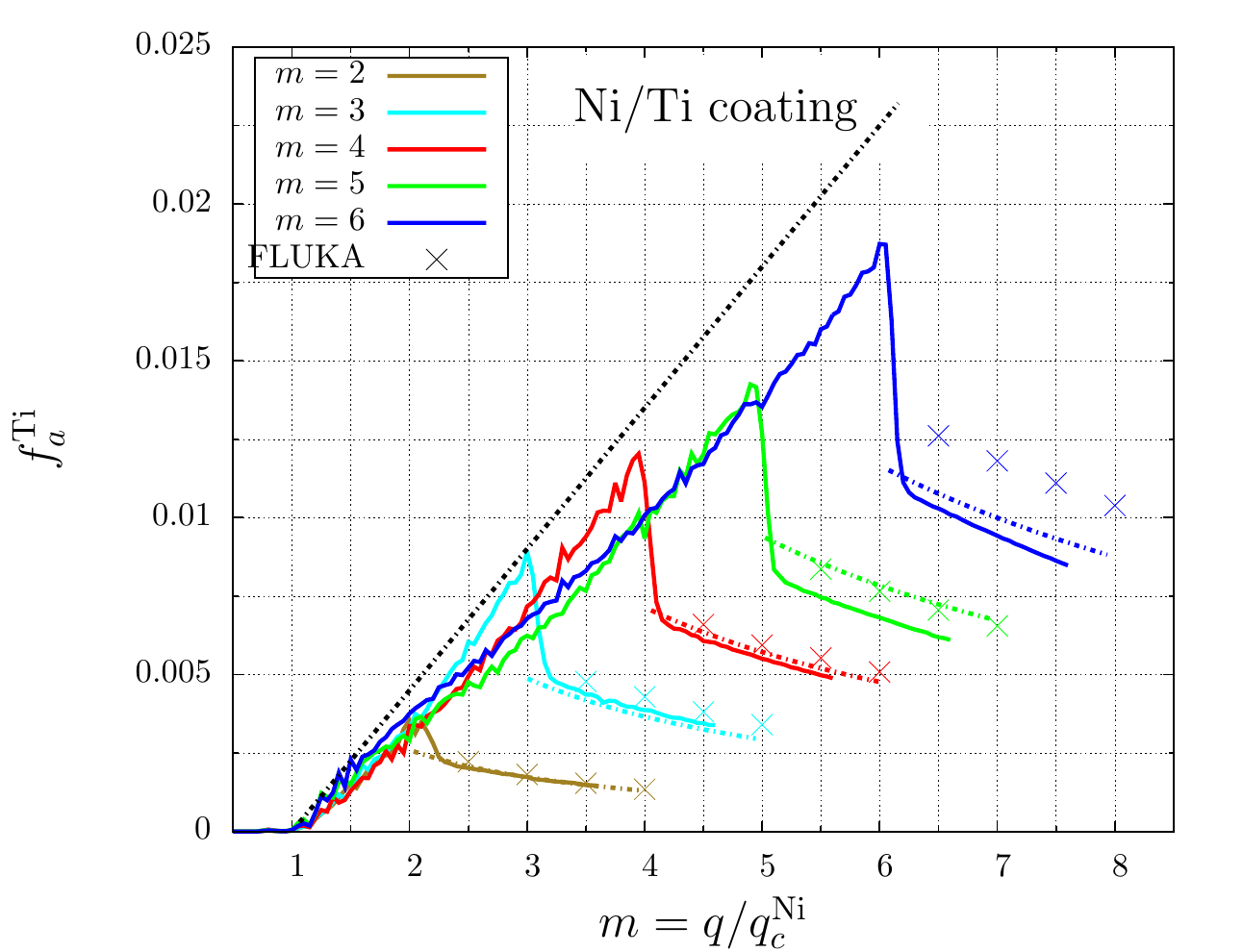}
  	
  	\caption{The radiative absorption by Ni and Ti per incident neutron are shown for supermirror composed of NiMo/Ti (left) and Ni/Ti (right). The calculations at wavelength $\lambda=5$~\AA \ (solid lines) are compared with the parameterization (dash-dotted lines) and, where it is possible, FLUKA transport code simulations.}
  	\label{fig:capture_all}
  \end{figure}

The calculation presented in the present work is thus beyond the capabilities of the transport Monte-Carlo codes in their present form. A direct comparison between our approach and the Monte-Carlo codes is only possible in a situation where the momentum transfer for a specular reflection is above the cutoff of the supermirror, that is when the reflection is essentially absent and all the neutrons are transmitted through the multilayer. To benchmark our calculation in this transmission regime with neutron transport simulation we used the FLUKA transport code \cite{battistoni2015overview,ferrari2005fluka}. The supermirror was approximated in the simulation as two parallel layers of Ni and Ti with thicknesses according to the total amount of the corresponding materials in the coating.  Results of the simulation for the  absorption probability of the NiTi coating per incident neutron are presented together with results of the theoretical calculation in Fig.~\ref{fig:capture_all}. The agreement is reasonable thus proving the validity of our approach.

\subsection{Parametrized absorption probabilities}
 
In this section simple guidelines for optimizing the gamma shielding of neutron guides equipped with supermirror are provided. We distinguish three distinct regimes A, B, and C for momentum transfers, where the radiative absorption probability per incident neutron can be expressed via simple universal functions of $m\equiv q/q_c^{\rm Ni}$.
 
 \subsection*{\underline{A: $ q \le q_c^{\rm Ni}$:}} The absorption probability per incident neutron is wavelength dependent. If $R_0$ is the low-angle reflectivity of the coating, the absorption per incident neutron is given by $(1-R_0)$ times the wavelength dependent probabilities from table~\ref{tab:captfracinf}. To be on the safe side, we suggest using values calculated for zero roughness.
 
\subsection*{\underline{B: $q_c^{\rm Ni} < q \le (m_{max}+0.1) \cdot q_c^{\rm Ni} $ :}} Below the cutoff of the coating (labeled as $m_{\max}$), the absorption probability per incident neutron is linear in $q$ and approximately  independent of wavelength.
Performing a fit to the linear parts of the curves $f_a^{\rm C}(q/q_c^{\rm Ni})$ for $m_{\max}=2\ldots 6$ coatings taken altogether gives $f_a^{\rm Ni}(m)=0.00407\cdot m$ for nickel, $f_a^{\rm Mo}(m)=0.00022\cdot m$ for molybdenum and
 $f_a^{\rm Ti}(m)=0.004\cdot m - 0.0045$ for titanium with $R^2=0.99$.
 To be on the safe side for shielding applications we suggest using a parameterization with an increased slope that slightly overshoots the calculated absorption probabilities:
 \begin{eqnarray}
 f_a^{\rm Ni} (m) &=& 0.005+ 0.005\cdot (m-1) \\ 
 f_a^{\rm Mo} (m) &=& 0.00027 + 0.00027\cdot (m-1)\\
 f_a^{\rm Ti} (m) &=& 0.0045\cdot (m-1)
 \end{eqnarray}
 
 \subsection*{\underline{C: $(m_{max}+0.1) \cdot q_c^{\rm Ni} < q$:}} 
 
 Beyond the cutoff of the coating, the approximately linear growth changes to a $f_a \propto 1/m$ behavior
 \begin{eqnarray}
 f_a^{\rm Ni}(m) &=& \cfrac{0.0025\cdot(m_{max}+0.1)^2}{m}\\
 f_a^{\rm Mo}(m) &=& \cfrac{0.000135\cdot(m_{max}+0.1)^2}{m}\\
 f_a^{\rm Ti}(m) &=& \cfrac{0.00225\cdot(m_{max}-0.9)(m_0+0.1)}{m}
 \end{eqnarray}
in accordance with Eq. \eqref{absbeyond}. We anticipate that the absorption drops by a factor of 2 from the value given by the linear parameterization anticipated in the regime B at $q/q_c^{\rm Ni} = m_{max}+0.1$. 

The calculated absorption probabilities per incident neutron of Ni and Ti for $2 \le m \le 6$ together with parameterized curves calculated for a wavelength $\lambda = 5$~\AA\  are plotted in Fig.~\ref{fig:capture_all}. The absorption probability for Mo in the NiMo/Ti coatings is 0.054 times the value for Ni and is not shown. As one can see from the figure, the suggested parameterization works with an accuracy of approximately 10\% for both Ni/Ti and NiMo/Ti supermirrors, which is sufficient for the use in shielding calculations. The difference between the theoretical and Monte-Carlo calculations at high $m$-values may be attributed to the effect of roughness which in the theoretical calculation effectively leads to a certain attenuation in the neutron wave before it reaches the reflection depth and hence lowers the overall probability for absorption.

\section{Conclusions.}
\label{sect:conclusions}

Our study shows  that the absorption probability of supermirror per incident neutron as calculated in a rigorous quantum-mechanical approach is essentially independent of the wavelength and follows a simple universal behavior as a function of the normal component of the incident neutron momentum $q_\perp$. For the absorption probability per non reflected neutron on the contrary such a universal behavior was not observed. It was also found that the absorption probability per non reflected neutron has a pronounced wavelength dependence due to a reduced reflectivity of supermirrors for shorter wavelengths.

We have derived a parameterization of absorption probabilities per incident neutron in the materials from which supermirror coatings are made of which is the key result of our paper. For a normal momentum component of the incident neutron exceeding the critical value for nickel, the absorption probability is expressed in a universal way via the normal momentum component and $m$-value of the coating independent of the wavelength. This allows for a straightforward evaluation of prompt gamma radiation along the supermirror coated neutron guides once the divergence profile and the spectrum of the transported beam are known.

We believe that implementing the parameterizations of the absorption rate suggested in section~\ref{sect:guidelines} in neutron-optical ray-tracing packages will provide an effective and simple to use tool for calculating dose rates and shielding requirements for the supermirror coated guides transporting high fluxes of neutrons.

\section*{Acknowledgment}
The authors acknowledge fruitful discussions with Dr. Tatsuhiko Ogawa on the current implementation of the supermirror physics in the PHITS particle transport code.

\end{document}